\documentclass[twoside]{iopart}

\usepackage{iopams,setstack}  
\usepackage{epsfig,bbold,psfrag,times}

\def\picdirectory{.}

\def\Name#1{#1}
\let\au\Name
\def\Review#1#2#3#4{#3 \emph{#1} \textbf{#2} #4}
\def\z#1#2#3#4{#4 \emph{#1} \textbf{#2} #3}

\makeatletter
\def\Book{\@ifnextchar[\@@Book\@Book}
\def\@Book#1#2#3#4{#4 \textit{#1} (#3: #2)}
\def\@@Book[#1]#2#3#4#5{#5 \textit{#2} #1 (#4: #3)}

\renewenvironment{abstract}{%
      \vspace{16pt plus3pt minus3pt}
      \begin{indented} \normalsize                    
      \item[]{\bfseries \abstractname.}\quad\rm\ignorespaces} 
      {\end{indented}\if@titlepage\newpage\else\vspace{18\p@ plus18\p@}\fi}

\renewcommand{\fulltable}[1]{\begin{table}
  \caption{#1}
  \lineup \footnotesize
  \begin{tabular*}{\textwidth}{@{}l*{15}{@{\extracolsep{0pt plus 12pt}}l}}}
\def\endfulltable{\end{tabular*}\end{table}}

\def\cite{\@ifnextchar[{\@tempswatrue\@citex}{\@tempswafalse\@citex[]}}

\def\@citex[#1]#2{%
  \if@filesw\immediate\write\@auxout{\string\citation{#2}}\fi%
  \leavevmode\unskip\ \@cite{\@collapse{#2}}{#1}}%

\def\@cite#1#2{[{#1\if@tempswa , #2\fi}]} %

\def\@collapse#1{%
{%
\let\@temp\relax
\@tempcntb\@MM
\def\@citea{}%
\@for \@citeb:=#1\do{%
\@ifundefined{b@\@citeb}%
{\@temp\@citea{\bf ?}%
\@tempcntb\@MM\let\@temp\relax
\@warning{Citation `\@citeb ' on page \thepage\space undefined}%
}%
{\@tempcnta\@tempcntb \advance\@tempcnta\@ne
\edef\MyTemp{\csname b@\@citeb\endcsname}%
\def\@tempa{\@temptokena=\bgroup}%
\afterassignment\@tempa %
\@tempcntb=0\MyTemp\relax}%
\ifnum\@tempcntb=0\relax%
\@tempcntb=\@MM
\@citea\MyTemp
\let\@temp = \relax
\else %
\edef\@tempd{\number\@tempcntb}%
\ifnum\@tempcnta=\@tempcntb %
\ifx\@temp\relax %
\edef\@temp{\@citea\@tempd}%
\else
\edef\@temp{\hbox{--}\@tempd}%
\fi
\else %
\@temp\@citea\@tempd
\let\@temp\relax
\fi
\fi
}%
\def\@citea{,$\,$}%
}%
\@temp %
}%
}%
\makeatother
\begin{document}

\topical{Dynamics of gelling liquids: a short survey}

\author[Henning L\"owe, Peter M\"uller, Annette Zippelius]{Henning L\"owe,
  Peter M\"uller and Annette Zippelius} 

\address{Institut f\"ur Theoretische Physik,
  Georg-August-Universit\"at, D--37077 G\"ottingen, Germany}

\def\today{21 December 2004}

\vspace{10pt plus 3pt minus 3pt}
\begin{indented} \normalsize
\item[] Version of \today \par\bigskip
\item[] \emph{Dedicated to Lothar Sch\"afer on the occasion of his
  60$^{\mathrm{th}}$ birthday}\par
\end{indented}

\begin{abstract}
  The dynamics of randomly crosslinked liquids is addressed via a Rouse- and a
  Zimm-type model with crosslink statistics taken either from bond percolation
  or Erd\H{o}s--R\'enyi random graphs. While the Rouse-type model isolates the
  effects of the random connectivity on the dynamics of molecular clusters,
  the Zimm-type model also accounts for hydrodynamic interactions on a
  preaveraged level. The incoherent intermediate scattering function is
  computed in thermal equilibrium, its critical behaviour near the sol-gel
  transition is analysed and related to the scaling of cluster diffusion
  constants at the critical point. Second, non-equilibrium dynamics is studied
  by looking at stress relaxation in a simple shear flow. Anomalous stress
  relaxation and critical rheological properties are derived. Some of the
  results contradict long-standing scaling arguments, which are shown to be
  flawed by inconsistencies.
\end{abstract}

%
\section{Introduction}
%

Gelling liquids are part of everyday life. One encounters them, for example,
when preparing a chocolate pudding or when sticking two materials together
with the help of glue. From a microscopic point of view, gelling liquids
consist of irregularly structured clusters of molecules or macromolecules. The
formation of these clusters is either a result of intermolecular association,
produced by e.g.\ van der Waals forces, electrostatic attractions or hydrogen
bonding, or a result of chemical reactions such as polycondensation,
polymerisation or vulcanisation induced by a chemical crosslinker 
\cite{Flo92,Lar99}.  Intermolecular association, also called physical
gelation, leads to 
weakly bound clusters, which typically form and dissolve reversibly in the
course of time during an experiment. On the other hand, chemical gelation
leads to permanent clusters at temperatures of interest, and it is this
situation that we will exclusively consider here. 

When increasing the concentration of crosslinks in a liquid (sol) one observes
a more and more viscous behaviour under shear stresses, until a sudden
transformation to an amorphous solid state takes place at a certain critical
crosslink concentration. This point marks the gelation transition or sol-gel
transition.  The static shear viscosity diverges at the transition, and the
onset of a static shear modulus is found.

Carothers \cite{Car36} was the first to interpret the gelation transition as
due to the formation of a macroscopic cluster of molecules in the system.  His
considerations were quantified and refined by Flory \cite{Flo41,Flo42} and
Stockmayer \cite{Sto43,Sto44} to what is nowadays called ``classical
theory'', a percolation model of tree-like structures, closely related to
percolation on Bethe lattices \cite{FiEs61}. So the classical theory arises
\cite{Ste77} in the mean-field approximation of lattice-bond percolation
\cite{StAh94}. Stauffer \cite{Sta76} and de Gennes \cite{Gen76} suggested the
latter as a mathematical model for gelation, in particular, if caused by
polycondensation.  Lattice-bond-percolation clusters may also contain loops,
and the spatial dimension becomes relevant, too. More importantly,
upon identifying the gelation transition with the lattice-bond-percolation
transition, it is revealed to be a continuous phase transition. Its driving
parameter is crosslink concentration, not temperature. Within this
theoretical picture, the critical behaviour at the gelation transition is
dictated by scaling and universality \cite{PlBe94,StAh94}.

The resulting predictions for static properties of gelation clusters agree
well with experiments in the vicinity of the sol-gel transition 
\cite{StCo82,Gen93}---a substantial improvement over the mean-field like
classical 
theory. As far as dynamical phenomena are concerned, a variety of competing
attempts have been made to seek an interpretation in terms of the percolation
picture, see e.g.\ \cite{Gen78,ArSa90,Sah94} for contradictory predictions
concerning the shear viscosity.  Yet, all of these attempts rely on more or
less \emph{ad hoc} assumptions needed to compensate for the lack of thermal
fluctuations or any sort of dynamics in a pure percolation model.  Rather, the
appropriate strategy should be to start from a (semi-) microscopic dynamical
model for gelation clusters, from which the desired link to quantities in
percolation theory can be \emph{deduced}. This route will be followed here.
Other analytical approaches to gelation from a microscopic model include 
\cite{GoCa96,ZiGo97,ThZi97,CaGo98,PeCa98,CaGo99,PeGo00,BrWe02,MuGo04}.
Among others, they describe thermostatic fluctuations in the gel phase and
calculate the static shear modulus. Computer simulations of microscopic models
for gelation have been done by e.g.\
\cite{GaAr00,VePl01,Jes02,GaAr02,PlVe03,JePl03,GaFi04}.

In this survey we will concentrate on the sol phase and report on results
obtained in \cite{BrGo97,BrLo99,BrLo01a,BrLo01b,BrMu02,Mul03,KuLo03,LoMu04}.
The dynamics of the sol phase is characterised by strong precursors of the
gelation transition, even well below it. These include anomalous,
stretched-exponential decays in time of both dynamical density correlations
\cite{MaWi88} and shear-stress relaxation \cite{WiMo97}. Both decays are
characterised by typical time scales which diverge when the critical crosslink
concentration is approached. Our exact results on critical rheological
properties contradict long-standing scaling arguments, which are shown to be
flawed by inconsistencies.

The paper is organised as follows. In Section~\ref{model} we briefly lay out a
suitable generalisation of the usual Rouse and Zimm model for linear polymers
to describe gelling liquids. The model is then used to investigate
time-dependent density fluctuations in Section~\ref{densfluc}.
Section~\ref{stressrelax} deals with stress relaxation and critical
rheological properties in a simple shear flow. Both Section~\ref{densfluc} and
Section~\ref{stressrelax} are subdivided in a part pertaining to the Rouse
model, a part pertaining to the Zimm model and a part where the results are
discussed and put in a wider perspective. Finally, Section~\ref{closing} adds
some closing remarks.

%
\section{Rouse and Zimm model for randomly crosslinked monomers}
\label{model}
%

In this section we give a brief description of a model which is to be
considered a theoretical minimal model for the dynamics of gelling complex
fluids. This model is a generalisation of one of the most fundamental models
of polymer physics \cite{Rou53,Zim56,BiCu87,DoEd88} to the case of randomly
connected monomers. In this context, it has been discussed before by e.g.\
\cite{Eic80,Cat84,NeEi85,Vil88,ShEi89,SoSc93,SoBl95,ZiKi96,BrGo97,%
  BrLo01a,BrLo01b,BrMu02,BlJu02,FeBl02,JuKo03,Mul03,KuLo03,LoMu04}. 

%
\subsection{Dynamical equation}
\label{dyneq}
%

We consider $N$ point-like monomers, which are characterised by their
time-dependent position vectors $\bi{R}_i(t)$, $i=1,\ldots, N$, in
three-dimensional Euclidean space $\mathbb{R}^{3}$. Permanently formed,
harmonic crosslinks connect $M$ randomly chosen pairs of particles $(i_e,
j_e)$, where $1 \le i_{e} \neq j_{e} \le N$ for all $e=1,\ldots,M$.  The
potential energy associated with these entropic Hookean springs takes the form
\begin{equation} 
  \label{energy} 
  V := \frac{3}{2a^2}\:\sum_{e=1}^M 
  \bigl( \bi{R}_{i_e}-\bi{R}_{j_e} \bigr)^2 
  =: \frac{3}{2a^2}\: \sum_{i,j=1}^{N}\bi{R}_{i}\cdot
  {\Gamma}_{i,j}\,\bi{R}_{j}\,,
\end{equation}
where the length $a>0$ plays the role of an inverse crosslink strength, and
physical units have been chosen such that $k_{\mathrm{B}}T = 1$. It will be
convenient to specify a given crosslink configuration
$\mathcal{G}:=\{(i_e,j_e)\}_{e=1}^M$ in terms of its $N\times N$-connectivity
matrix $\Gamma$, which is defined by the right equality in (\ref{energy}).
For part of what follows this setting could be generalised to the crosslinking
of $N$ identical molecular units which consist themselves of a given number of
monomers that are connected in some fixed manner, such as $N$ identical
chains, rings or stars of monomers \cite{BrLo01a,BrLo01b}.  For the ease of
presentation, however, we will not consider such a generalisation here.

We study the dynamics of these harmonically crosslinked monomers in the
presence of an incompressible solvent fluid, which may induce hydrodynamic
interactions between them. Hydrodynamic interactions will be incorporated on a
preaveraged level in the spirit of Kirkwood and Riseman \cite{KiRi48} and Zimm
\cite{Zim56}. This is a traditionally accepted way of doing so albeit the
limitations of this approach are still not sufficiently well explored
\cite{BiCu87,DoEd88}.  We also allow for the presence of an externally
imposed, simple shear flow in $x$-direction 
\begin{equation}
  \label{shearflow}
  \mathbf{v}(\bi{r},t) :=   \dot{\gamma}(t) y \, \bi{e}_{x}
\end{equation}
with a time-dependent shear rate $\dot{\gamma}(t)$. Here $\bi{r}=(x,y,z)$.
A purely
relaxational monomer dynamics is then described by \cite{BiCu87,DoEd88}
\begin{equation} 
  \label{zimm}
  \frac{{\rmd}}{\rmd t} \bi{R}_{i}(t) -
  \mathbf{v} \bigl(\bi{R}_{i}(t),t \bigr)   = 
  -\sum_{j=1}^{N} \mathsf{H}^{\mathrm{eq}}_{i,j}\:
  \frac{\partial V}{\partial \bi{R}_{j}(t)} 
  + \boldsymbol{\xi}_{i}(t) 
\end{equation}
for $i=1,\ldots,n$. This is the defining equation of the \emph{Zimm model
  for crosslinked monomers (in solution)}. The rest of this subsection is
devoted to a brief explanation and discussion of (\ref{zimm}), see
\cite{KuLo03,LoMu04} for more details.

The jointly Gaussian thermal noises $\boldsymbol{\xi}_{i}$ in (\ref{zimm})
have zero mean and covariance $ \overline{\boldsymbol{\xi}_{i}(t)
  \,\boldsymbol{\xi}_{j}^{\dagger}(t')} = 2\,\mathsf{H}^{\mathrm{eq}}_{i,j}
\,\delta(t-t') \boldsymbol{\mathsf{1}}$, as is required by the
fluctuation-response theorem. As usual, the $\boldsymbol{\xi}_{i}$
``thermalize'' the system in the long-time
limit.  Here, the dagger denotes the transposition of a vector, $\delta$ the
Dirac-delta function and $\boldsymbol{\mathsf{1}}$ the $3\times 3$-unit
matrix.

Interactions between the monomers and the solvent fluid are subsumed in the
spatially isotropic and homogeneous preaveraged mobility matrix
\begin{equation} 
  \label{preav}
  \mathsf{H}^{\mathrm{eq}}_{i,j} :=
  \frac{1}{\zeta}\;\Bigl[ \delta_{i,j}
  +(1-\delta_{i,j})\;
  h\!\left(\kappa^2\,\pi/\mathcal{R}_{i,j} \right)\Bigr].
\end{equation}
It emerges \cite{KuLo03,LoMu04} from taking Oseen's expression
\cite{Ose10,KiRi48} for the mobility tensor and averaging it with respect to
the suitably normalised Boltzmann weight $\sim \rme^{-V}$.  However, when it is
indispensable to have a positive definite mobility matrix in the sequel, we
will replace the Oseen tensor with the Rotne--Prager--Yamakawa tensor
\cite{RoPr69,Yam70} in this procedure.  Depending on which tensor is used, the
function $h$ in (\ref{preav}) is given by \cite{Fix83}
\begin{equation}
  \label{haa}
  \fl
  h(x) := \left\{ 
    \begin{array}{c@{\qquad\quad}l}
      \sqrt{x/\pi} & \mbox{Oseen}, \\
      \mathrm{erf}(\sqrt{x})- (1- \rme^{-x})/ \sqrt{\pi x}
      & \mbox{Rotne--Prager--Yamakawa}.
    \end{array}
\right.
\end{equation}
The expression in the second line of (\ref{haa}) involves the error function
$\mathrm{erf}$ and reduces to the expression of the Oseen case asymptotically
as $x \downarrow 0$. The diagonal term in the preaveraged mobility
matrix(\ref{preav}), which is proportional to the Kronecker symbol
$\delta_{i,j}$, accounts for a frictional force with friction constant $\zeta$
that acts when a monomer moves relative to the externally imposed flow field
(\ref{shearflow}). The non-diagonal term reflects the solvent-mediated average
influence of the motion of monomer $j$ on monomer $i$. The parameter $\kappa
:= \sqrt{6/\pi}\,\zeta/(6\pi\eta_{s}a)$ involves the solvent viscosity
$\eta_{s}$ and serves as the coupling constant of the hydrodynamic
interaction.  Formally setting $\kappa=0$ in (\ref{preav}) yields
$\mathsf{H}^{\mathrm{eq}}_{i,j} = \zeta^{-1} \delta_{i,j}$, and the Zimm model
for crosslinked monomers reduces to the \emph{Rouse model for
  crosslinked monomers} \cite{BrGo97,BrLo99,BrLo01a,BrLo01b,BrMu02,Mul03}
\begin{equation} 
  \label{rouseeq}
  \frac{{\rmd}}{\rmd t} \bi{R}_{i}(t) -
  \mathbf{v} \bigl(\bi{R}_{i}(t),t \bigr)   = 
  - \frac{1}{\zeta} \frac{\partial V}{\partial \bi{R}_{i}(t)} 
  + \boldsymbol{\xi}_{i}(t) \,,
\end{equation}
where $i=1,\ldots,n$ and the jointly Gaussian thermal noises
$\boldsymbol{\xi}_{i}$ have zero mean and covariance $
\overline{\boldsymbol{\xi}_{i}(t) \,\boldsymbol{\xi}_{j}^{\dagger}(t')} =
(2/\zeta)\,\delta(t-t') \boldsymbol{\mathsf{1}}$. It is only for convenience
that we introduced the Rouse model as the special case $\kappa=0$ of the Zimm
model here. Physically, it has its own standing as \emph{the} minimal model
for polymer melts under theta conditions, see e.g.\ \cite{BiCu87,DoEd88} for
the case of linear polymer chains. In particular, all the approximations that
entered the derivation of the (off-diagonal part of the) preaveraged mobility
matrix $\mathsf{H}^{\mathrm{eq}}$ do not affect the Rouse model, of course. 

It remains to explain the quantity $\mathcal{R}_{i,j}$ in (\ref{preav}), which
is simply the mean squared displacement between monomers $i$ and $j$ in the
thermal-equilibrium state characterised by the suitably normalised Boltzmann
weight $\sim \rme^{-V}$. In order to write down a formula for
$\mathcal{R}_{i,j}$, let us remark that, by construction, the connectivity
matrix $\Gamma \equiv \Gamma(\mathcal{G})$ is block-diagonal with respect to
the clusters of a given crosslink configuration $\mathcal{G}$ (which are the
maximal connected components of $\mathcal{G}$). Moreover,
$\Gamma(\mathcal{G})$ possesses as many zero eigenvalues as there are clusters
in $\mathcal{G}$. This is easily seen from the fact that the centre of mass of
each cluster does not feel a force from the potential energy $V$.  Hence,
$\Gamma$ cannot be inverted, but it possesses a Moore--Penrose pseudo-inverse
$\mathsf{Z}$ \cite{Al72}, which is the inverse of $\Gamma$ on the complement
of its zero eigenspace and zero elsewhere. It can be represented as
$\mathsf{Z} := (\mathsf{1}-\mathsf{E}_0)/{\Gamma}$, where $\mathsf{E}_{0}$
denotes the projector on the zero eigenspace of $\Gamma$ in $\mathbb{R}^{N}$
and $\mathsf{1}$ denotes the $N\times N$-unit matrix. The mean-squared
displacement $\mathcal{R}_{i,j}$ is then given in terms of $\mathsf{Z}$
according to
\begin{equation}
  \label{resist}
  \fl
  \mathcal{R}_{i,j} := \left\{ \begin{array}{c@{\qquad\quad}l}
\mathsf{Z}_{i,i}+\mathsf{Z}_{j,j}-2\mathsf{Z}_{i,j} & \mbox{if $i$ and $j$
  belong to the same cluster,} \\[.5mm]
+\infty & \mbox{otherwise.}\end{array}\right.
\end{equation}
There is also another interpretation for $\mathcal{R}_{i,j}$, which we will
use below: Viewing each monomer as an electric contact and each crosslink as a
unit Ohmian resistor connecting two contacts, $\mathcal{R}_{i,j}$ is the
effective electric resistance between the contacts $i$ and $j$ of this
corresponding electrical resistor network \cite{KlRa93}. This \emph{exact}
correspondence between Hookean bead-spring clusters and Ohmian electrical
resistor networks relies on the linearity of Hooke's and Ohm's law.

Since both the connectivity matrix $\Gamma$ and the preaveraged mobility
matrix $\mathsf{H}^{\mathrm{eq}}$ are block-diagonal, it follows that clusters
move \emph{independently} of each other in this model.  The salient feature of
the Zimm and Rouse equations (\ref{zimm}) and (\ref{rouseeq}) is that they are
linear in the monomers' positions. Hence, they admit an explicitly known
solution. The results we present in this paper rely heavily on this solution.

%
\subsection{Average over crosslink ensemble}
\label{average}
%

So far, everything in this section was meant for an arbitrary but fixed
realisation $\mathcal{G}$ of $M$ crosslinks among $N$ monomers. For practical
reasons, $\mathcal{G}$ can never be determined experimentally in
macroscopically large gelling fluids.  Neither should physically meaningful
observables depend on specific microscopic details of $\mathcal{G}$, but only
on some macroscopic characteristics of it. Therefore, we follow the general
philosophy of the theory of disordered systems and take $\mathcal{G}$ as an
element of a statistical ensemble of crosslink configurations, within which it
occurs with probability $P_{N}(\mathcal{G})$. The just made statement on
physically meaningful observables $A(\mathcal{G})$ now translates into a
\emph{self-averaging property}: the two quantities $A(\mathcal{G})$ and its
ensemble average $\sum_{\mathcal{G}'} P_{N}(\mathcal{G}') A(\mathcal{G}')$
coincide (with probability one) in the macroscopic limit.  Therefore we will
compute the macroscopic limit
\begin{equation}
  \langle A\rangle := 
  \lim_{N\to\infty} \sum_{\mathcal{G}} P_{N}(\mathcal{G}) A(\mathcal{G})
\end{equation}
of such averages with a fixed \emph{crosslink concentration} $c :=
\lim_{N\to\infty} M/N$. This will be
done for two different
crosslink ensembles.\\
\indent (i)~~Clusters are generated according to three-dimensional continuum
percolation, which is closely related to the intuitive picture of gelation,
where monomers are more likely to be crosslinked when they are close to each
other. Since continuum percolation and lattice percolation are believed to be
in the same universality class \cite{StAh94}, we employ the scaling
description of the latter. It predicts \cite{StAh94} a cluster-size
distribution of the form
\begin{equation}
  \label{clustersizedist}
  \tau_n  \sim n^{-\tau} \exp\{-n/n^{*}\}
\end{equation}
for $\varepsilon:=(c_{\mathrm{crit}}-c) \ll 1$ and $n\to\infty$ with a typical
cluster size $n^{*}(\varepsilon) \sim \varepsilon^{-1/\sigma}$ that diverges
as $\varepsilon \to 0$. Here, $\sigma$ and $\tau$ are
(static) critical exponents, see Table~\ref{percexp} below for their numerical
values.\\
\indent (ii)~~Each pair of monomers is chosen independently with equal
probability $c/N$, corresponding to Erd\H{o}s--R\'enyi random graphs, which
are known to resemble the critical properties of mean-field percolation
\cite{Ste77}. After performing the macroscopic limit, there is no macroscopic
cluster for $c<c_{\mathrm{crit}}=1/2$ and almost all clusters are trees
\cite{ErRe60}.  Furthermore, all $n^{n-2}$ trees of a given ``size'' $n$, that
is, with $n$ monomers, are equally likely.  The cluster-size distribution can
also be cast into the scaling form (\ref{clustersizedist}) with the exactly
known critical exponents $\tau$ and $\sigma$ listed below in
Table~\ref{percexp}.

%
\section{Time-dependent density fluctuations}
\label{densfluc}
%

In this section we address dynamical properties of gelling liquids in thermal
equilibrium. Therefore we will assume throughout this section that there is no
externally imposed shear flow, i.e.\ $\dot{\gamma}=0$.

Experiments \cite{MaWi88,MaWi91} on quasi-elastic light
scattering in gelling liquids allow to measure how spatial density
fluctuations of a given wave vector $\bi{q}$ are correlated to each other at
different times $t$. This information is encoded in the incoherent
intermediate scattering function
\begin{equation}
  \label{Sdef}
  S(\bi{q},t) :=  \lim_{t_{0}\to -\infty}
  \overline{\frac{1}{N} \sum_{i=1}^N
    \rme^{\,\rmi\,\bi{q}\cdot [\bi{R}_{i}(t+t_{0}) -
      \bi{R}_{i}(t_{0})]}}\;.
\end{equation}
The right-hand side of (\ref{Sdef}) is determined by the solution
$\bi{R}_{i}(t)$ of the dynamical equation (\ref{zimm})
for a given crosslink realisation $\mathcal{G}$ and with initial conditions
being imposed at time $t_{0}$.  The average over the thermal noise and the
subsequent limit $t_{0} \to -\infty$ in (\ref{Sdef}) ensure that the system
reaches its thermal-equilibrium state. Then, for large retardation times $t$,
one expects \cite{Gen79,MaWi88} that this correlation is determined by the
slowest relaxation processes in the system. Due to the independent motion of
different clusters in the model under consideration, the slowest relaxation
processes correspond to the centre-of-mass diffusion of whole clusters of
monomers. This argument can be quantified---see e.g.\ \cite{BrGo97},
\cite{KuLo03} or Eq.\ (4.12) in \cite{BrLo01a}---and yields
\begin{equation}
  \label{Stcluster}
  S(\bi{q},t)   \;\;\stackrel{t\to\infty}{\sim} \;\;
  \sum_{k=1}^{K} \frac{N_{k}}{N}\, \exp\{- q^{2} t 
  D(\mathcal{N}_{k})\} \,.
\end{equation}
Here we have set $q:=|\bi{q}|$ and introduced the clusters $\mathcal{N}_{k}$,
$k=1,\ldots,K$, of the given crosslink configuration $\mathcal{G}$. The number
of monomers in the cluster $\mathcal{N}_{k}$ is denoted by $N_{k}$ and
\begin{equation}
  \label{clusterdiff}
  \fl
  D({\cal N}_k):=\lim_{t\to\infty}  
  \frac{1}{6t}\; \overline{\bigl[\bi{R}_{\mathrm{CM}_{\,k}}(t) -
    \bi{R}_{\mathrm{CM}_{\,k}}(0)\bigr]^2}   
  =\bigg( \sum_{i,j\in\mathcal{N}_k}
  \left[\frac{1}{\mathsf{H}^{\mathrm{eq}}}\right]_{i,j}\bigg)^{-1} 
\end{equation}
defines its diffusion constant in terms of the mean-square displacement of its
centre of mass $\bi{R}_{\mathrm{CM}_{\,k}}(t) := N_{k}^{-1}
\sum_{i\in\mathcal{N}_{k}}\bi{R}_{i}(t)$. The right equality in
(\ref{clusterdiff}) follows from a short calculation with the exact solution
of the dynamical equation (\ref{zimm}). It was
previously established in \cite{Oet87}.  Another diffusion constant has been
introduced by Kirkwood \cite{DoEd88,BiCu87}
\begin{equation} \label{kirkwood}
 \widehat{D}(\mathcal{N}_k) := 
 \frac{1}{N_k^2}\sum_{i,j\in
   \mathcal{N}_k}\mathsf{H}^{\mathrm{eq}}_{i,j} \,.
\end{equation}
It provides an upper bound to the former, 
\begin{equation}
  \label{jepe}
  D(\mathcal{N}_k)\leq \widehat{D}(\mathcal{N}_k)  \,,
\end{equation}
as can be shown by applying the Jensen--Peierls inequality, see e.g. Sect.\ 8c
in \cite{Sim79}, to (\ref{clusterdiff}).  Customarily, one also defines an
effective diffusion constant $D_{\mathrm{eff}}$ for the whole gelling liquid
by
\begin{equation}
  \label{Deff}
  D_{\mathrm{eff}}^{-1} := \lim_{q\to 0} q^{2}
  \int_{0}^{\infty} \rmd t\,S(\bi{q},t) 
  = \sum_{k=1}^{K} \frac{N_{k}}{N}\; \frac{1}{D(\mathcal{N}_{k})} \;.  
\end{equation}
Since $S(\bi{q},t)$ is expected to develop a time-persistent part in the gel
phase, $D_{\mathrm{eff}}$ is expected to vanish when approaching the gelation
transition from the sol side.

%
\subsection{Rouse dynamics}
%

We recall from Sect.\ \ref{dyneq} that in the absence of hydrodynamic
interactions, $\kappa =0$, we have $\mathsf{H}^{\mathrm{eq}}_{i,j} =
\zeta^{-1} \delta_{i,j}$. Hence, the cluster-diffusion constant
(\ref{clusterdiff}) and the Kirkwood diffusion constant (\ref{kirkwood}) are
equal 
\begin{equation}
  \label{rousediff}
  D(\mathcal{N}_{k}) = \widehat{D}(\mathcal{N}_{k}) = \frac{1}{\zeta N_{k}}\;, 
\end{equation}
and inversely proportional to the number of monomers in the cluster
\cite{BrGo97}. In other words, cluster topology does not influence diffusion
within Rouse dynamics.

Next, we discuss the long-time behaviour of the incoherent intermediate
scattering function in the macroscopic limit. According to
Sect.~\ref{average}, this amounts to calculating the average of
(\ref{Stcluster}) 
\begin{equation}
  \langle S(\bi{q},t)\rangle
  \stackrel{t\to\infty}{\sim} \biggl\langle \sum_{k=1}^{K}
  \frac{N_{k}}{N}\, \exp\{- q^{2} t D(\mathcal{N}_{k})\} \biggr\rangle \,.
\end{equation}
Thanks to (\ref{rousediff}) this average is easily performed by reordering the
clusters according to their size
\begin{equation}
  \label{2step}
  \langle S(\bi{q},t) \rangle  \stackrel{t\to\infty}{\sim}
  \sum_{n=1}^{\infty} n \tau_{n} \, \e^{-q^{2}t/(\zeta n)} \,,
\end{equation}
where 
\begin{equation}
  \tau_{n} := \biggl\langle \sum_{k=1}^{K}
  \frac{1}{N}\, \delta_{N_{k}, n} \biggr\rangle 
\end{equation}
is the cluster-size distribution and (\ref{2step}) holds in the absence of an
infinite cluster. Using the scaling form (\ref{clustersizedist}) of
$\tau_{n}$, we find \cite{BrGo97,BrLo01a}
\begin{equation}
  \label{Sasym}
  \fl
  \langle S(\bi{q},t) \rangle \stackrel{t\to\infty}{\sim}
  \biggl(\frac{\zeta}{q^{2} 
  t}\biggr)^{y} \;
  \left\{ 
    \begin{array}{c@{\qquad}l}
     1   & \varepsilon =0 \,, \\[0.5ex]
     \relax [t/t_{q}^{*}(\varepsilon)]^{(y-1/2)/2} \exp\{ - \mathrm{const.}\;
     [t/t_{q}^{*}(\varepsilon)]^{1/2} \}  & \varepsilon >0 \,.
    \end{array}
\right.
\end{equation}
At the critical point, the long-time decay is algebraic with a critical
exponent $y=\tau -2$. In the sol phase one has a Kohlrausch or
stretched-exponential behaviour with a time scale that diverges as
$t_{q}^{*}(\varepsilon) \sim (\zeta /q^{2}) \varepsilon^{-\mu}$ with a
critical exponent $\mu = 1/\sigma$, when the critical point is approached.

For the effective diffusion constant (\ref{Deff}) we conclude from
(\ref{Sasym}) that it vanishes like
\begin{equation}
  \langle D_{\mathrm{eff}} \rangle \sim \lim_{q \downarrow 0} \bigl[ q^{2}
  t_{q}(\varepsilon) \bigr]^{-(3 -\tau)} \sim \varepsilon^{a} \qquad\quad
  \mbox{with~~} a=(3-\tau)/\sigma
\end{equation}
as $\varepsilon\downarrow 0$. 

The exponent $a$ could have also been deduced
directly from the right expression in (\ref{Deff}). Indeed, given any
\emph{cluster-additive observable}, a reordering of the clusters according to
their size yields 
\begin{equation}
  \label{reordering}
  \langle A \rangle 
  = \biggl\langle
  \sum_{k=1}^{K} \frac{N_{k}}{N}\; A(\mathcal{N}_{k}) \biggr\rangle  
  = \sum_{n=1}^{\infty} n\tau_{n}\langle A \rangle_{n} \,,
\end{equation}
where
\begin{equation}
  \langle A \rangle_{n} := \frac{1}{\tau_{n}} \; \biggl\langle
  \sum_{k=1}^{K}  \frac{1}{N}\, \delta_{N_{k}, n} \,
  A(\mathcal{N}_{k})\biggr\rangle  
\end{equation}
is the partial average of $A$ over all clusters of a given size $n$. Now, if
the partial averages exhibit the critical divergence
\begin{equation}
  \label{partialav}
  A_{n} :=  \langle A\rangle_{n}\big|_{\varepsilon =0}\sim n^b\,,
\end{equation}
then 
\begin{equation}
  \label{critscaling}
  \langle A \rangle  \sim\varepsilon^{-u}
  \quad\mbox{as}\quad\varepsilon\downarrow 0 \quad\mbox{with}\quad
  u=(2-\tau+b)/\sigma ,
\end{equation}
provided that $u>0$.

%
\subsection{Zimm dynamics}
%

\begin{table}[tb]
  \caption{\label{percexp} Numerical values for the critical exponents of the
    cluster-size distribution (\ref{clustersizedist}) and the two fractal
    dimensions of Gaussian phantom clusters in (\ref{gausshaus}).
    The values are listed for cluster statistics according to
    three-dimensional bond percolation ({3D}) and Erd\H{o}s--R\'enyi
    random graphs ({ER}).}
  \begin{indented}
  \item[]
    \begin{tabular}{@{}lllll}
      \br
         & $\tau$  & $\sigma$ & $d_{s}$  &  $d_{f}^{(\mathrm{G})}$ \\ \mr
         \lineup 
      3D & 2.18 & 0.45  &  1.33 &  3.97 \\
      ER & 5/2  & 1/2   &  4/3  &  4 \\
      \br
    \end{tabular}
  \end{indented}
\end{table}

In contrast to the free-draining limit described by Rouse dynamics in the last
subsection, one expects that with hydrodynamic interactions being present,
cluster topology will have an influence on the diffusion constants. 

For simplicity, let us start with the Kirkwood diffusion constant.  In order
to extract a size dependence out of $\widehat{D}$, we look at the average
$\langle \widehat{D}\rangle_{n}$ over all clusters of a given size $n$ and
study its behaviour as a function of $n$.  More specifically, we will perform
this average precisely at the critical concentration $c_{\mathrm{crit}}$,
where we expect an algebraic decrease as $n\to\infty$ due to the absence of
any other length scale at criticality. Indeed, using the Oseen tensor for the
hydrodynamic interactions we deduce from (\ref{kirkwood}), (\ref{preav}) and
(\ref{haa}) that
\begin{equation}
  \label{kirkcrossover}
  \fl
  \widehat{D}_{n} := \langle\widehat{D}\rangle_{n}\big|_{c=c_{\mathrm{crit}}} 
  = \frac{1}{\zeta n} + \frac{\kappa}{\zeta n^{2}} \, \sum_{\substack{i,j=1
  \crcr  i\neq j}}^{n} \langle
  \mathcal{R}_{i,j}^{-1/2}\rangle_{n}\big|_{c=c_{\mathrm{crit}}}  \stackrel{n
  \to\infty}{\sim} \frac{1}{\zeta}\left(\frac{1}{n} +
  \frac{\lambda \kappa}{n^{1/d_{f}^{(\mathrm{G})}}} \right)  \;,
\end{equation}
where $\lambda$ is some dimensionless proportionality constant. The asymptotic
behaviour of the average over the resistances in (\ref{kirkcrossover}) is
derived in \cite{KuLo03}. The derivation has to distinguish between the two
different cases for the crosslink ensemble. For Erd\H{o}s--R\'enyi random
graphs the asymptotics can be deduced from the exact probability distribution
of $\mathcal{R}_{i,j}$ in \cite{MeMo70}. For three-dimensional bond
percolation we use the scaling form of the probability distribution, which was
established within two-loop order of a renormalisation-group treatment of an
associated field theory \cite{HaLu87,StJa99}. Equation (\ref{kirkcrossover})
involves the fractal Hausdorff dimension
\begin{equation}
  \label{gausshaus}
  d_{f}^{(\mathrm{G})} := 2d_{s} /(2-d_{s})
\end{equation}
of Gaussian phantom clusters, which also determines the scaling of their radius
of gyration according to \cite{Cat84,Vil88,SoBl95}
\begin{equation}
  \label{gyr}
  R_{\mathrm{gyr},n} := \biggl[ \frac{1}{2n^{2}} \sum_{i,j=1}^{n} 
  \bigl\langle (\bi{R}_{i} -\bi{R}_{j})^{2} \bigr\rangle_{n}^{\phantom{2}}
  \big|_{c=c_{\mathrm{crit}}} \biggr]^{1/2} 
  \stackrel{n \to\infty}{\sim} n^{1/d_{f}^{(\mathrm{G})}}  .
\end{equation}
The other fractal dimension in (\ref{gausshaus}) is the \emph{spectral
  dimension} $d_{s}$ of the incipient percolating cluster 
\cite{NaYa94,BuHa96}. Their numerical values are listed in Table~\ref{percexp}.
We conclude from (\ref{kirkcrossover}) that
$\widehat{D}_{n}$ shows a crossover from Rouse behaviour $\widehat{D}_{n} \sim
n^{-1}$ for $n < \widehat{n}(\kappa) \sim
\kappa^{-1/(1-1/d_{f}^{(\mathrm{G})})}$ to Zimm behaviour
\begin{equation}
  \label{bhat}
  \widehat{D}_{n} \sim n^{-1/d_{f}^{(\mathrm{G})}} \sim  1/
  R_{\mathrm{gyr},n}
\end{equation}
for asymptotically large $n > \widehat{n}(\kappa)$.

\begin{figure}[t]
  \epsfig{file=\picdirectory/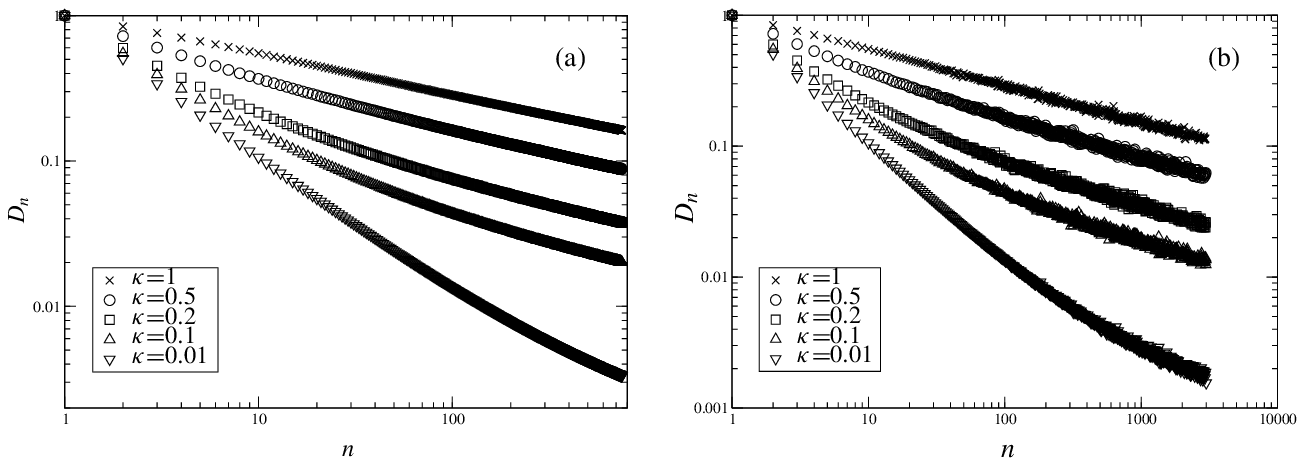, clip=,
    width=\textwidth}    
  \vspace{-3ex}
  \caption{(a) $D_{n}$ at the gel point for mean field percolation and
    different hydrodynamic interaction strengths.
    (b) Same for three-dimensional bond percolation.
    \label{fig:1}}   
  \par\vspace{3ex}
  \epsfig{file=\picdirectory/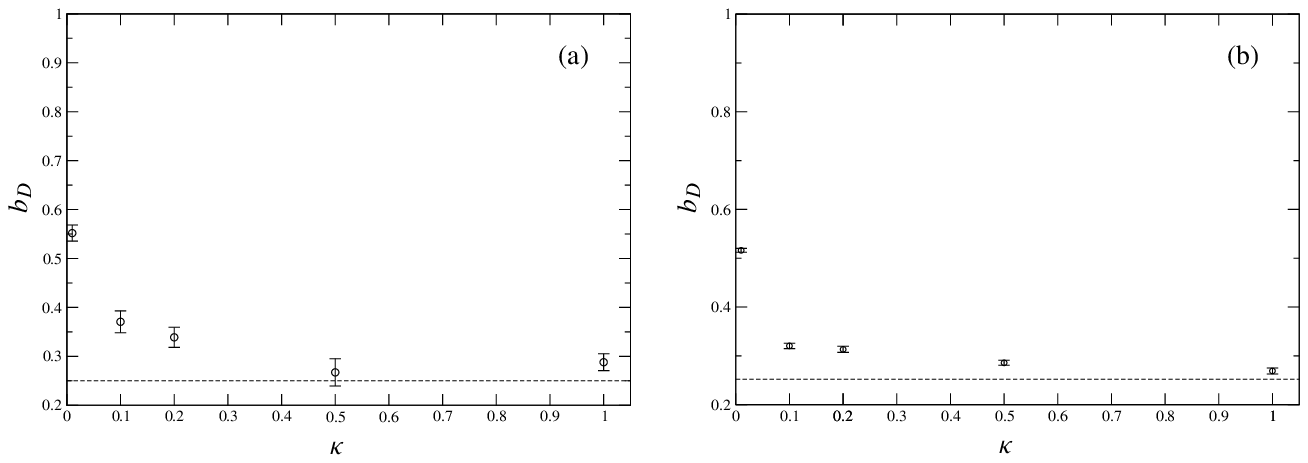, clip=,
    width=\textwidth}    
  \vspace{-3ex}
  \caption{(a) Critical exponents $b_{D}$, corresponding to a power-law
    fit $D_{n} \sim n^{-b_{D}}$ in Fig.~\ref{fig:1}(a). \linebreak[2]
    (b) Same for Fig.~\ref{fig:1}(b).
    \label{fig:2}}
\end{figure}

Now we turn to the averaged diffusion constant 
\begin{equation}
  \label{dnscale}
  D_{n} :=  \langle D\rangle_{n}\big|_{c=c_{\mathrm{crit}}}
  \stackrel{n\to\infty}{\sim}  n^{-b_{D}}
\end{equation}
of clusters of size $n$ at the gel point, which is also expected to obey a
critical scaling for large cluster sizes $n$. From the Jensen-Peierls
inequality $D_{n} \le \widehat{D}_{n}$, see (\ref{jepe}), we then infer the
inequality
\begin{equation}
  b_{D} \ge 1/d_{f}^{(\mathrm{G})}
\end{equation}
for the critical exponents. Figure~\ref{fig:1} shows numerical data for the
cluster diffusion constant $D_{n}$, plotted against $n$, for different values
of the hydrodynamic interaction strength $\kappa$. The crosslink ensemble in
Fig.~\ref{fig:1}(a) corresponds to Erd\H{o}s--R\'enyi random graphs. In
Fig.~\ref{fig:1}(b) crosslinks were chosen according to three-dimensional
bond-percolation. In the numerical computations we have used
$\mathsf{H}^{\mathrm{eq}}$ corresponding to the Rotne--Prager--Yamakawa tensor
so that a positive definite mobility matrix is always guaranteed.  Like the
Kirkwood diffusion constant, $D_{n}$ also exhibits a crossover from Rouse to
Zimm behaviour at a cluster size comparable to $\widehat{n}({\kappa})$.
Figure~\ref{fig:2} shows the exponent $b_{D}$ of the power-law fit
(\ref{dnscale}) to the data of Fig.~\ref{fig:1} in the large $n$-regime for
the different values of $\kappa$. The horizontal dashed lines in
Figs.~\ref{fig:2}(a) and~(b) correspond to the exponent value
$1/d_{f}^{(\mathrm{G})}$ of the respective Kirkwood diffusion constant. The
bigger exponent values that occur for small values of $\kappa$ still show
residual Rouse behaviour for the largest system sizes we treated. For bigger
values of $\kappa$ the crossover can hardly be felt any more in the largest
systems, and the extracted exponent value $b_{D}$ corresponds to Zimm
dynamics. This value is very close to the scaling exponent in (\ref{bhat}) for
the Kirkwood diffusion constant, and, in fact, we conjecture that
\begin{equation}
  \label{DexpZ}
  b_{D}= 1/d_{f}^{(\mathrm{G})}.
\end{equation}

We now turn to the long-time behaviour of the incoherent intermediate
scattering function (\ref{Sdef}). The asymptotics (\ref{Stcluster}),
(\ref{reordering}) and Jensen's inequality yield the lower bound \cite{KuLo03}
\begin{equation}
  \label{SboundZ}
  \langle S(\bi{q},t)\rangle \ge \sum_{n=1}^{\infty} n\tau_{n} \e^{-q^{2} t
  D_{n}} \,. 
\end{equation}
In fact, there is numerical evidence that this inequality actually captures
the correct long-time asymptotics of $\langle S(\bi{q},t)\rangle$. Evaluating
the 
right-hand side of (\ref{SboundZ}) for large times $t$, this then leads to the
scaling form \cite{KuLo03}\footnote{Note that there is a misprint in the
  second line after Eq.\ (30) in \cite{KuLo03}. The algebraic prefactor in the
  scaling form of the function $s(\lambda)$ should read $\lambda^{x(y-1/2)}$
  instead of $\lambda^{xy}$.}
\begin{equation}
  \label{SasymZ}
  \fl
  \langle S(\bi{q},t)\rangle \stackrel{t\to\infty}{\sim}
  \biggl(\frac{\zeta}{q^{2} 
  t}\biggr)^{y} \;
  \left\{ 
    \begin{array}{c@{\qquad}l}
     1   & \varepsilon =0 \,, \\[0.5ex]
     \relax [t/t_{q}^{*}(\varepsilon)]^{x(y-1/2)} \exp\{ - \mathrm{const.}\;
     [t/t_{q}^{*}(\varepsilon)]^{x} \}  & \varepsilon >0 
    \end{array}
\right.
\end{equation}
with the time scale $t_{q}^{*}(\varepsilon) \stackrel{\varepsilon \downarrow
  0}{\sim} q^{-2} \varepsilon^{-z}$. The exponents are given by 
\begin{equation}
  x = (1+b_{D})^{-1}\,, \qquad y= (\tau -2)/b_{D} \,, \qquad   z= b_{D}/\sigma
\end{equation}
and are expressed in terms of $b_{D}\approx 0.25$, see (\ref{DexpZ}) and
Table~\ref{percexp}. The Rouse 
limit (\ref{Sasym}) of (\ref{SasymZ}) corresponds to setting $b_{D}=1$ in the
above expressions. 

The critical vanishing
\begin{equation}
  \langle D_{\mathrm{eff}}\rangle \sim \varepsilon^{a}
  \hspace{1cm}\mathrm{with} \hspace{1cm} a=(2-\tau
  +b_{D})/\sigma 
\end{equation}
of the effective diffusion constant follows from directly from
(\ref{reordering}) -- (\ref{critscaling}) provided that $a>0$. This condition
is fulfilled for three-dimensional bond percolation where $a\approx 0.16$, but
violated for Erd\H{o}s--R\'enyi random graphs.  Finally, we like to point out
that, regardless of the cluster statistics, the ensemble averaged diffusion
constant $\langle D \rangle$ never vanishes at the critical point. This is
simply because it has non-vanishing contributions from all clusters, which add
up.

%
\subsection{Discussion} \label{densitydiscuss}
%

We have studied the critical scaling $D_{n} \sim n^{-b_{D}}$ of the averaged
cluster diffusion constants over clusters of size $n$ and used it to obtain
the scaling behaviour of the intermediate incoherent scattering function
$\langle S(\bi{q},t)\rangle$ near criticality. The associated critical
exponents are 
summarised in Table~\ref{tab:1}. Within Rouse dynamics cluster diffusion
constants are inversely proportional to the cluster size $n$, irrespective of
the cluster topology, that is, $b_{D}=1$. Zimm dynamics leads to $b_{D}=
1/d_{f}^{(\mathrm{G})}$, see (\ref{DexpZ}), and topology does play a role:
Indeed, it is well known \cite{DoEd88} that within Zimm dynamics the diffusion
constant of a \emph{linear chain} of $n$ monomers decreases as $n^{-1/2}$.
Since $b_{D} \approx 0.25 < 1/2$, this means that, on average, a monomer in a
branched cluster feels less friction---which is intuitively appealing, because
monomers in the interior of a cluster should be dragged along.  Second,
(\ref{gyr}), (\ref{dnscale}) and (\ref{DexpZ}) imply for Zimm dynamics that
$D_n \sim 1/R_{\mathrm{gyr},n}$. Hence, this relation does not only hold for
linear chains, 
for which it has been well known \cite{DoEd88}, but in an average sense for
\emph{all} percolation clusters.

Concerning the scaling exponents of the incoherent intermediate scattering
function, Table~\ref{tab:1} shows that neither Rouse nor Zimm dynamics
provides even a reasonably good description of the experimental findings,
despite their strong scatter. There are several reasons for the discrepancies
between the model predictions and experiments. ~(i)~ Our results pertain to
$\theta$-conditions, in so far as ex\-cluded-volume interactions have been
neglected in the models.  Excluded-volume interactions could cause a swelling
of the clusters, which results in a different fractal Hausdorff dimension.
~(ii)~ We chose cluster statistics according to three-dimensional bond
percolation. This accounts well for crosslinking in a dense melt, say, but not
in dilute solutions. ~(iii)~ It has been suggested \cite{MaWi91} that
hydrodynamic interactions between monomers in a cluster are screened by
smaller clusters in the reaction bath so that the Rouse rather than the Zimm
model should apply. Our analysis supports this conclusion in so far as the
exponents of the Rouse model are closer to the experimental values. So the
more striking failure of the Zimm model can be traced back to a too slow decay
of $D_{n}$ with $n$. ~(iv)~ Preaveraging of the hydrodynamic interactions is
an uncontrolled approximation, and it remains to be seen what a full treatment
of hydrodynamic interactions predicts for the critical dynamics of gelling
solutions.

\begin{table}[tb]
  \caption{\label{tab:1} Summary of critical
    exponents for cluster diffusion constants and the incoherent intermediate
    scattering function (see Eqs.\ (\ref{dnscale}), (\ref{Sasym}) and
    (\ref{SasymZ}) for their definitions). The  
    numerical values for Rouse and Zimm dynamics---listed for
    cluster statistics from three-dimensional bond percolation (3D) and
    Erd\H{o}s--R\'enyi random graphs (ER)---are
    compared to experimental findings.}
  \begin{indented}
  \item[]
    \begin{tabular}{@{}llllllll}
      \br
               & \centre{2}{Zimm} & \centre{2}{Rouse} &&& \\ \ns\ns
               & \crule{2} & \crule{2} &&& \\
      Exponent & 3D & ER & 3D & ER  & \cite{MaWi91} & \cite{AdDe88} &
      \cite{BaBu92} \\ \mr \lineup
      $b_{D}$ & 0.25 & 1/4      & 1    & 1   &      &            &      \\
      $x$     & 0.80 & 4/5      & 1/2  & 1/2 & 0.66 & 0.3 -- 0.8 & 0.64 \\
      $y$     & 0.71 & 2        & 0.18 & 1/2 & 0.27 & 0.2 -- 0.3 & 0.34 \\
      $z$     & 0.56 & 1/2      & 2.22 & 2   & 2.5  &            &      \\
      $a$     & 0.16 & $^{(*)}$ & 1.82 & 1   & 1.9  & 0.5 -- 1    & 1.9 \\
      \br
    \end{tabular}
  \item[] $^{(*)}$ no divergence
  \end{indented}
\end{table}

%
\section{Stress relaxation}
\label{stressrelax}
%

Gelling liquids exhibit striking rheological properties which have been
continuously studied over the years by experiments
\cite{AdDe81,DuDe87,MaAd88,AdMa90,DeBo93,CoGi93,VlCh98,ToFa01},  
theories \cite{Sta76,Gen78,Cat84,MaAd89,RuZu90,BrGo97,ZiGr98,%
  BrLo99,BrLo01a,BrAs01,BrLo01b,BrMu02} and simulations
\cite{GaAr00,VePl01,GaAr02,ShAd02,PlVe03,JePl03,GaFi04}. 
For example, when subjected to the homogeneous shear flow (\ref{shearflow}),
distinct relaxation patterns are observed, which are due to the participation
of many different excitation modes of all sorts of clusters. More precisely,
experiments suggest the scaling form \cite{DaLa90,WiMo97,%
  DuDe87,MaAd88,AdMa90,ToFa01,MaAd89}
\begin{equation}
  \label{gscale}
   \langle G(t) \rangle \sim t^{-\Delta} g(t/\overline{t}) \hspace{1.5cm}
  \mathrm{with}\hspace{1.5cm} 
  \overline{t}(\varepsilon) \sim \varepsilon^{-\overline{z}}  
\end{equation}
for the macroscopic (shear-) stress-relaxation function in the sol phase for
asymptotically long times $t$ and crosslink concentrations close to the
critical point, \emph{i.e.}\ for $\varepsilon \ll 1$.  The typical relaxation
time $\overline{t}$ diverges with a critical exponent $\overline{z}>0$ for
$\varepsilon\downarrow 0$. The scaling function $g$ is of order unity for
small arguments so that one finds the algebraic decay $ \langle G(t)
\rangle\sim t^{-\Delta}$ with a critical exponent $0<\Delta\le 1$ for
$t\to\infty$ at the critical point. For large arguments, $g$ decreases faster
than any inverse power. Sometimes a stretched exponential has been proposed
for $g$ in this asymptotic regime \cite{AdMa90,MaAd89}.

In this section we will investigate to what extent such critical properties
can be predicted by the Rouse and the Zimm model.  Thus we will explore the
consequences of the dynamics (\ref{zimm}), resp.\ (\ref{rouseeq}), in the
presence of the externally applied simple shear flow (\ref{shearflow}). In
reaction to the flow, the system of crosslinked monomers builds up an
intrinsic shear stress.  Following Kirkwood, see e.g.\ Chap.~3 in
\cite{DoEd88} or Chap.\ 16.3 in \cite{BiCu87}, this shear stress is given by
the force per unit area exerted by the monomers
\begin{equation}
  \label{stress}
  \boldsymbol{\sigma}(t) = \lim_{t_{0}\to-\infty}
  -\frac{\rho_{0}}{N}\sum_{i=1}^{N} 
  \overline{\bi{F}_{i}(t) \bi{R}^{\dagger}_{i}(t)}.
\end{equation}
Here, $\bi{R}_{i}(t)$ is the solution of the equation of motion (\ref{zimm})
with some initial condition at time $t_{0}$ in the distant past (so that the
noise average yields a thermalized state in which all transient effects
stemming from the initial condition have died out).  Moreover, $\rho_{0}$
stands for the monomer concentration and $\bi{F}_{i}(t):=-\partial V/\partial
\bi{R}_i(t)$ is the net spring force acting on monomer $i$ at time $t$.  
The explicit computation \cite{BrLo01a,BrLo01b} of the right-hand side of
(\ref{stress}) yields
\begin{equation}
  \label{linearresponse}
  \boldsymbol{\sigma}(t) =G(0) \,
  \boldsymbol{\mathsf{1}} + 
  \int_{-\infty}^t {\rmd} t' \; G(t-t')\,
  \dot{\gamma}(t') 
  \left(
    \begin{array}{ccc}
      2\,\int_{t'}^{t}{\rmd}s\:\dot{\gamma}(s) & 1 & 0\\
      1&0&0\\
      0&0&0\\
    \end{array}
  \right)
\end{equation}
for \emph{arbitrary} strengths of the shear rate $\dot{\gamma}(t)$. Here, we
have defined the stress-relaxation function 
\begin{equation}
  \label{stressrelaxation}
  G(t) := 
  \frac{\rho_{0}}{N}\; {\rm Tr}
  \biggl[(\mathsf{1}-\widetilde{\mathsf{E}}_0)
  \exp\biggl(-\,\frac{6t}{a^{2}}\;  
  \widetilde{{\Gamma}}\biggr)\biggr]
\end{equation}
as a trace over the matrix exponential of $\widetilde{{\Gamma}} :=
(\mathsf{H}^{\mathrm{eq}})^{1/2} {\Gamma}\, (\mathsf{H}^{\mathrm{eq}})^{1/2}$.
Due to the occurrence of the spectral projector $\widetilde{\mathsf{E}}_0$ on
the kernel of $\widetilde{{\Gamma}}$, this trace is effectively restricted to
the subspace of non-zero eigenvalues.

For a time-independent shear rate $\dot{\gamma}$, the shear stress
(\ref{linearresponse}) is also independent of time. The viscosity $\eta$ is
then related to shear stress via
\begin{equation}
  \label{viscositydef}
  \eta  :=\frac{\sigma_{x,y}}{\dot{\gamma}\rho_{0}} =
  \frac{1}{\rho_0}\int_0^{\infty}{\rmd}t\: G(t)=
  \frac{a^2}{3}\,
  \frac{1}{2N}
  \:{\rm Tr}\!\left[\frac{\mathsf{1}-\widetilde{\mathsf{E}}_{0}}%
    {\rule{0pt}{2.3ex}\widetilde{{\Gamma}}}\right] \,.
\end{equation}
Apparently, the viscosity is determined by the trace of the Moore--Penrose
inverse of $\widetilde{\Gamma}$.  The normal stress coefficients
are given by
\begin{equation}
  \label{normalstress}
  \Psi^{(1)} :=\frac{\sigma_{x,x} - \sigma_{y,y}}{\dot{\gamma}^2\rho_0} 
  = \frac{2}{\rho_0}\int_0^{\infty}\!\!{\rmd}t\;t G(t)=
  \left(\frac{a^{2}}{3}\right)^2
  \frac{1}{2N}
  \:{\rm Tr}\!\left[\frac{\mathsf{1}-\widetilde{\mathsf{E}}_0}%
   {\widetilde{\Gamma}^{2}}\right]
\end{equation}
and
\begin{equation}
  \label{normstressdef}
  \Psi^{(2)} :  =\frac{\sigma_{y,y} - \sigma_{z,z}}{\dot{\gamma}^2\rho_0} =0
  \,.  
\end{equation}
The vanishing of $\Psi^{(2)}$ is typical for Rouse/Zimm-type models and 
has been well known for the case of linear polymers
\cite{DoEd88}. Since $\widetilde{{\Gamma}}$ is block-diagonal with
respect to the clusters, the observables $G(t)$,
$\eta$ and $ \Psi^{(1)}$ are all cluster-additive in
the sense of (\ref{reordering}).  

The scaling form (\ref{gscale}) of the macroscopic stress-relaxation function 
$ \langle G(t) \rangle$ implies that
the macroscopic viscosity and first normal stress coefficient exhibit a
critical divergence
\begin{equation}
  \label{critrheo}
  \langle \eta \rangle \sim \varepsilon^{-k}
  \hspace{1cm}\mathrm{and}\hspace{1cm}
  \langle \Psi^{(1)} \rangle \sim
  \varepsilon^{-\ell} 
\end{equation}
at the sol-gel transition as $\varepsilon\downarrow 0$
with critical exponents given by
the scaling relations \cite{WiMo97,BrMu02} 
\begin{equation}
  \label{scaling}
  k = z(1-\Delta) \hspace{1cm}\mathrm{and}\hspace{1cm}
  \ell = z(2-\Delta) = k+z\,.
\end{equation}
Thus, it suffices to know any two of the four critical exponents $\Delta,
z, k$ and $\ell$.

%
\subsection{Rouse dynamics}
%

For Rouse dynamics we have $\widetilde{\Gamma} = \Gamma /\zeta $ so that the
computation of the stress-relaxation function, the viscosity or the first
normal stress coefficient requires the knowledge of spectral properties of the
connectivity matrix $\Gamma$.

Concerning the macroscopic viscosity $\langle\eta\rangle$, there are several
ways of calculating the critical exponent $k$ in (\ref{critrheo}). The
different ways explore connections to problems in different branches of
research. Given a cluster $\mathcal{N}_{k}$, the trace of the Moore--Penrose
inverse of $\Gamma(\mathcal{N}_{k})$ can be expressed in terms of the
resistances (\ref{resist}) according to \cite{BrLo99,BrLo01a}
\begin{equation}
  \label{exactcorr}
  \eta(\mathcal{N}_{k}) = \frac{\zeta a^{2}}{6 N_{k}} \; \Tr \biggl[
  \frac{\mathsf{1} - \mathsf{E}_{0}
    (\mathcal{N}_{k})}{\Gamma(\mathcal{N}_{k})}\biggr] 
  =  \frac{\zeta a^{2}}{12 N_{k}^{2}} \, \sum_{i,j \in \mathcal{N}_{k}}
  \mathcal{R}_{i,j} \,.
\end{equation}
We stress that this is an exact relation \cite{KlRa93}. It has nothing to do
with electrical analogues put forward in scaling arguments \cite{Gen79}. For
the case of Erd\H{o}s--R\'enyi random graphs there are only tree clusters for
$c < c_{\mathrm{crit}} = 1/2$. In this special case the resistance
$\mathcal{R}_{i,j}$ reduces to the graph distance of $i$ and $j$ in
$\mathcal{N}_{k}$, and the right-hand side of (\ref{exactcorr}) is known as
the Wiener index ${W}(\mathcal{N}_{k})$ in graph theory. From a
graph-theoretical point of view, the right equality in (\ref{exactcorr})
follows also as an application of the matrix-tree theorem, see e.g.\
\cite{Mer94}, Thm.~5.5. Moreover, the partial averages $\langle{W}\rangle$ are
exactly known \cite{MeMo70}, and, using (\ref{reordering}), one finds the exact
result \cite{BrLo99,BrLo01a}
\begin{equation}
  \label{etaER}
  \langle\eta\rangle = \frac{\zeta a^{2}}{24 c} \, \biggl[ \ln \biggl(
  \frac{1}{1 -2c} \biggr) -2c \biggr]\,.
\end{equation}
It can be interpreted as a critical divergence with exponent $k=0$.
Alternatively, (\ref{etaER}) can also be
obtained from a replica approach \cite{BrLo01a} instead of using graph theory.
The replica approach is also capable of providing us with higher inverse
moments $\langle N^{-1} \Tr\, [(\mathsf{1} - \mathsf{E}_{0})/\Gamma^{\nu}]
\rangle$ for not too large positive integers $\nu$ \cite{BrAs01}. Using these
results for $\nu=2$, a (somewhat lengthy) exact expression for $\Psi^{(1)}$ was
derived in \cite{BrMu02} for crosslink statistics from Erd\H{o}s--R\'enyi
random graphs. It exhibits the critical behaviour
\begin{equation}
  \label{psiER}
  \langle \Psi^{(1)}\rangle  
  \sim \varepsilon^{-\ell} \qquad
  \mathrm{with}\qquad \ell =3  \,.
\end{equation}

Now we turn to the crosslink ensemble of three-dimensional bond percolation.
In order to proceed from (\ref{exactcorr}) in this case,
one needs to know the average resistance $\langle
\mathcal{R}_{i,j} \rangle_{n}$ between two nodes in bond-percolation clusters
of size $n$. Luckily, random electric resistance networks have been studied
extensively, and the asymptotic behaviour 
\begin{equation}
  \label{errn}
  \langle\mathcal{R}_{i,j} \rangle_{n} \sim n^{b_{\eta}} \qquad
  \mathrm{with}\qquad b_{\eta} := (2/d_{s}) -1 
\end{equation}
can be extracted \cite{BrLo99,BrLo01a} from highly developed
renormalisation-group treatments of an associated field theory 
\cite{HaLu87,StJa99}. Thus, (\ref{exactcorr}), (\ref{reordering}) and
(\ref{critscaling}) lead to the critical behaviour $\langle\eta\rangle \sim
\varepsilon^{-k}$ with
\begin{equation}
  \label{viscexp}
  k=  (1-\tau +2/d_{s})/\sigma
\end{equation}
as $\varepsilon\downarrow 0$. Of course, this exact scaling behaviour reduces
to the Erd\H{o}s--R\'enyi result $k=0$ from (\ref{etaER}), when inserting the
appropriate mean-field values for the exponents. 

None of the above approaches is able to yield any of the other critical
exponents $\Delta$ and $z$---or also $\ell$ in the case of three-dimensional
percolation statistics. Here, a connection to random walks in random
environments is helpful. For the time being, let us concentrate on the case of
three-dimensional percolation statistics, where the maximum number of bonds
emanating from any vertex is limited to $m=6$ on the simple cubic lattice.
Now, consider a random walker---coined ``blind ant'' by de~Gennes
\cite{Gen76}---that moves along a bond from one site to another in the same
cluster at discrete time steps \cite{StAh94,BuHa96,AlOr82,HaBe02}.  If the ant
happens to visit site $i$ at time $s$, which is connected with $m_{i} \le m$
bonds to other sites, then it will move with equal probability $1/m$ along any
one of the $m_{i}$ bonds within the next time step and stay at site $i$ with
probability $1 - m_{i}/m$. By definition of the connectivity matrix $\Gamma$
of the cluster, one has $\Gamma_{ii} = m_{i}$ for its diagonal matrix
elements, $\Gamma_{ij} = -1$ if two different sites $i \neq j$ are connected
by a bond and zero otherwise. Hence, the associated master equation for the
ant's sojourn probability $p_{i}(s)$ for site $i$ at time $s$ reads
\begin{equation}
  p_{i}(s+1) = (1- \Gamma_{ii}/m) p_{i}(s) + \sum_{j \neq i}
  (-\Gamma_{ij}/m) p_{j}(s)\,,
\end{equation}
which is equivalent to
\begin{equation}
  \label{master}
  p_{i}(s+1) -p_{i}(s) = - m^{-1} \sum_{j} \Gamma_{ij} p_{j}(s)\,.
\end{equation}
Here the
summation extends over all sites in the cluster. For long times $s
\gg 1$, it is legitimate to replace the difference (quotient) on the
left-hand side of (\ref{master}) by a derivative. This yields the
solution $p_{i}(s) = \bigl[\rme^{-s\Gamma/m}\bigr]_{ii_{0}}$, which
corresponds to the initial condition $p_{i}(0) = \delta_{i,i_{0}}$.
Next we consider $P^{(n)}(s) := \left.\langle p_{i_{0}}(s)
  \rangle_{n}\right|_{\varepsilon=0}$, the mean return probability to the
starting point after time $s$, where the average is taken over all critical
percolation clusters with $n$ sites. Clearly, these definitions are
independent of the starting point $i_{0}$, because on average there is no
distinguished site by assumption. Thus we can also write
\begin{equation}
  \label{nreturn}
  P^{(n)}(s) = \left. \left\langle \frac{1}{n} \; {\mathrm{Tr}}\;
    \rme^{-s\Gamma/m} \right\rangle_{n} \right|_{\varepsilon =0}
\end{equation}
for finite $n$. The return probability behaves as \cite{BuHa96,HaBe02,AlOr82}
\begin{equation}
  \label{nreturnas}
  P^{(n)}(s) \sim s^{-d_{s}/2} \mathcal{F}(s/s_{n}) +1/n\,,
\end{equation}
where $s_{n}\sim n^{2/d_{s}}$ and the cut-off function $\mathcal{F}(x)$ is of
order one for $x \lesssim 1$ and decreases rapidly to zero for $x
\to\infty$. Basically, (\ref{nreturnas}) says that for times $s \gg s_{n}$ the
walker has no memory of where he had started from. For times $s\lesssim
s_{n}$ the fractal-like nature of a cluster at $c=c_{\mathrm{crit}}$ leads
to an algebraic decrease of the return probability, which involves the spectral
dimension $d_{s}$. Now, assuming that $\langle G(t)\rangle$ obeys the scaling
form (\ref{gscale}), the information provided by (\ref{nreturnas}) for
$c=c_{\mathrm{crit}}$ is sufficient to conclude \cite{Mul03} the exponent
relations 
\begin{equation}
  \label{stressexp}
  \Delta = \frac{d_{s}}{2}\, (\tau-1) \qquad\mathrm{and}\qquad
  z = \frac{2}{d_{s}\sigma}\,.
\end{equation}
When plugging (\ref{stressexp}) into (\ref{scaling}), we recover
(\ref{viscexp}) and get the new scaling relation 
\begin{equation}
  \label{psiexp}
  \ell = (1-\tau +4/d_{s})/\sigma\,.
\end{equation}
Since, the critical behaviour of Erd\H{o}s--R\'enyi random graphs coincides
with that of mean-field percolation, we get the missing exponents $\Delta$ and
$z$ for that case by inserting the mean-field values into (\ref{stressexp}).

%
\subsection{Zimm dynamics}
%

The matrix $\widetilde{\Gamma}$, which determines stress relaxation, is by far
more complicated than $\Gamma$ in the presence of hydrodynamic interactions. In
particular, it reflects cluster topology only in a much more subtle way than
$\Gamma$. In fact, it was that apparent encoding of topology in
$\Gamma$ that made the analytical methods of the last subsection work. In the
absence of suitable analytical tools, numerical methods remain to investigate
stress relaxation in the Zimm model. 

We determined the scaling as $n\to\infty$ of the partial averages
\begin{equation}
  \label{zimmscaling}
  \eta_{n} := \langle \eta\rangle_{n}\big|_{\varepsilon=0} \sim
  n^{b_{\eta}}\qquad\mathrm{and}\qquad 
  \Psi^{(1)}_{n} := \langle \Psi^{(1)}\rangle_{n}\big|_{\varepsilon=0} \sim
  n^{b_{\Psi}} 
\end{equation}
at criticality by numerically diagonalising $\widetilde{\Gamma}$ and
performing the disorder average over the crosslink ensemble \cite{LoMu04}. The
critical exponents $k$ and $\ell$ then follow from (\ref{critscaling}). All
numerical computations were done with the Rotne--Prager--Yamakawa tensor for
the hydrodynamic interactions. The reader who is interested in more details of
the numerical computations is referred to \cite{LoMu04}.

\begin{figure}[t]
  {\leavevmode 
    \epsfig{file=\picdirectory/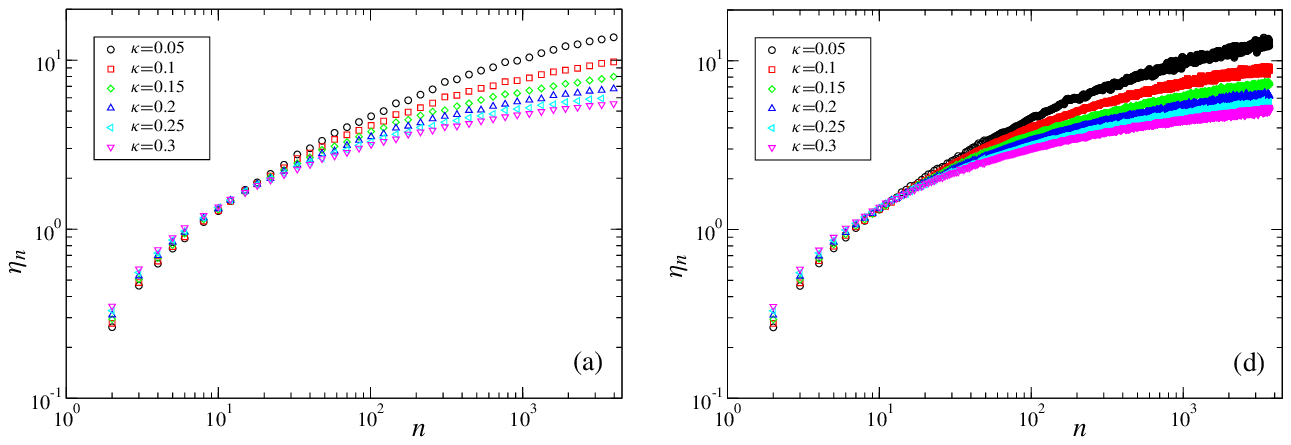, clip=,
      width=\textwidth}\par\medskip
    \epsfig{file=\picdirectory/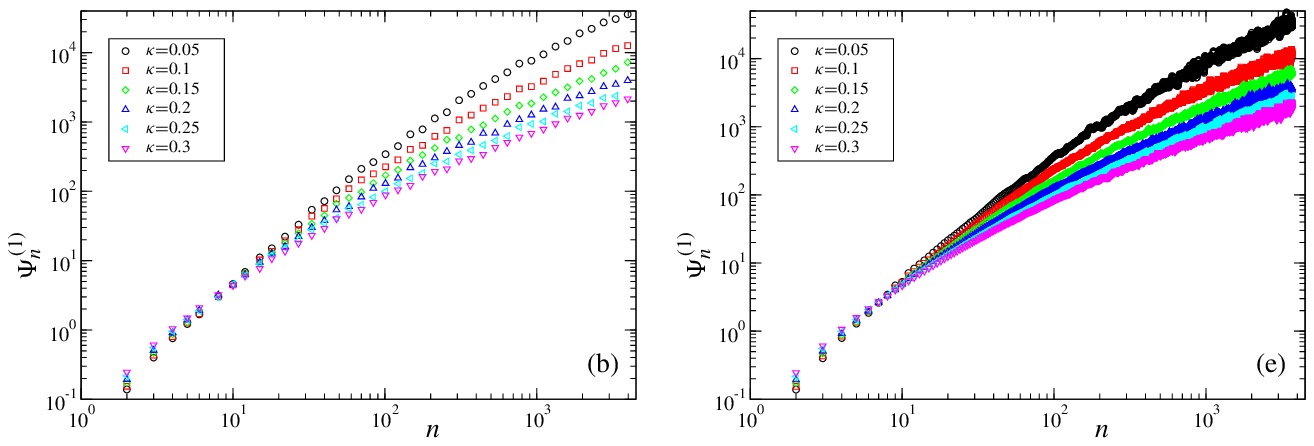, clip=,
      width=\textwidth}\par\medskip
    \epsfig{file=\picdirectory/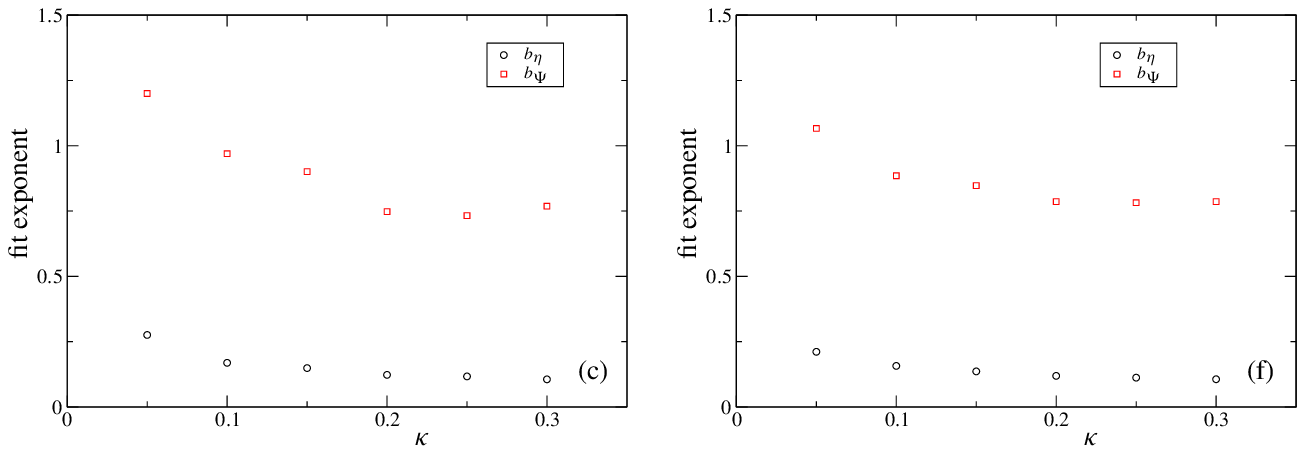, clip=,
      width=\textwidth}
    \caption{Numerical data to determine the scaling
      (\ref{zimmscaling}) for random clusters in the case of
      Erd\H{o}s--R\'enyi random graphs (left column) and
      three-dimensional bond percolation (right column). 
      In each case the averaged viscosity
      ${\eta}_n$ (top) and normal stress coefficient $\Psi^{(1)}_n$
      (middle) are plotted for different strengths of the hydrodynamic
      interaction parameter $\kappa$ as a function of the cluster size $n$
      on a double logarithmic scale. Power-law fits to the data
      yield the exponents $b_{\eta}$ and $b_{\Psi}$ as
      a function of $\kappa$ (bottom).
      \label{fig:3}
    }
    }
\end{figure}

In Figs.~\ref{fig:3}(a) and~(b) we plot ${\eta}_{n}$ and $\Psi^{(1)}_n$ as a
function of $n$ on a double-logarithmic scale for different values of the
hydrodynamic interaction parameter $\kappa$. Crosslink statistics were chosen
according to Erd\H{o}s--R\'enyi random graphs. The exponents $b_{\eta}$ and
$b_{\Psi}$ are obtained from power-law fits in the large $n$-range and are
displayed in Fig.~\ref{fig:3}(c). The viscosity exponent decreases from
$b_{\eta}=0.28$ for $\kappa=0.05$ to $b_{\eta}=0.11$ for $\kappa=0.3$. We
recall from (\ref{exactcorr}) and (\ref{errn}) that the Rouse exponent for
$\kappa=0$ is exactly given by $b_{\eta}=1/2$.  The exponent $b_{\Psi}$ of the
normal stress coefficient ranges from $b_{\Psi}=1.2$ for $\kappa=0.05$ to
$b_{\Psi}=0.73$ for $\kappa=0.25$.  The exact Rouse value $b_{\Psi}= (4/d_{s})
-1 =2$ for $\kappa=0$ follows from (\ref{psiER}) and (\ref{critscaling}).

The same is done for three-dimensional bond percolation in the right column of
Fig.~\ref{fig:3}. Figures~\ref{fig:3}(d) and~(e) contain $\eta_{n}$ and
$\Psi^{(1)}_{n}$, respectively, as a function of $n$ on a double-logarithmic
scale for different values of $\kappa$. The exponents $b_{\eta}$ and
$b_{\Psi}$, extracted by fitting the curves in Figs.~\ref{fig:3}(d) and~(e) to
a power law for large $n$, are shown in Fig.~\ref{fig:3}(f). The numerical
values for $b_{\eta}$ are nearly identical to those obtained for
Erd\H{o}s--R\'enyi random graphs.  Again, one observes a decrease from
$b_{\eta}=0.21$ for $\kappa=0.05$ to $b_{\eta}=0.11$ for $\kappa=0.3$. The
exponent $b_{\Psi}$ of the normal stress coefficient ranges from
$b_{\Psi}=1.1$ for $\kappa=0.05$ to $b_{\Psi}=0.78$ for $\kappa=0.25$. The
corresponding exact Rouse values $b_{\eta} = (2/d_{s}) -1 \approx 1/2$ and
$b_{\Psi} = (4/d_{s}) -1 \approx 2$ for $\kappa=0$ follow from
(\ref{viscexp}) and (\ref{psiexp}) in the last subsection together with
(\ref{critscaling}). 

A careful analysis of the data in \cite{LoMu04} reveals that the true Zimm
exponents $b_{\eta}$ and $b_{\Psi}$ are universal in $\kappa$ and that their
seeming dependence on $\kappa$ in Figs.~\ref{fig:3}(c) and~(f) is most likely
due to finite-size effects. More precisely, for small $\kappa$ the data suffer
from a crossover to their respective Rouse values so that they come out too
large. For large $\kappa$, on the other hand, the asymptotics $h(x) \sim 1 -
(\pi x)^{-1/2}$ as $x\to\infty$ of the lower line in (\ref{haa}) leads to a
slower growth of $b_{\eta}$ and $b_{\Psi}$ at intermediate $n$. Hence, the
exponents come out too small for larger $\kappa$.  The most reliable values
for the universal Zimm exponents $b_{\eta}$ and $b_{\Psi}$ should be obtained
from around $\kappa \approx 0.3$. It it these values which are listed in
Table~\ref{stresstab} below. The critical behaviour of the averaged viscosity
$\langle\eta\rangle \sim \varepsilon^{-k}$ and of the averaged first normal
stress coefficient $\langle\Psi^{(1)} \rangle\sim \varepsilon^{-\ell}$ for a
polydisperse gelling solution of crosslinked monomers then follows from
(\ref{critscaling}). For the viscosity this implies a {\it finite} value at
the gel point for both, Erd\H{o}s--R\'enyi random graphs and three-dimensional
bond percolation. In contrast, the first normal stress coefficient is found to
diverge with an exponent that depends on the cluster statistics. Choosing the
cluster statistics according to Erd\H{o}s--R\'enyi random graphs, we find
$\ell \approx 0.54$.  The case of three-dimensional bond percolation leads to
the higher value $\ell \approx 1.3$. These exponent values are less than a
third in magnitude than the corresponding exact analytical predictions of the
Rouse model from (\ref{psiexp}) with the corresponding cluster statistics. All
exponent values are summarised in Table~\ref{stresstab}.

%
\subsection{Discussion} \label{stressdiscuss}
%

\fulltable{\label{stresstab} 
    Summary of critical exponents for stress
    relaxation (see Eqs.\ (\ref{gscale}), (\ref{critrheo}) and
    (\ref{zimmscaling}) for their definitions). The numerical values for Rouse
    dynamics are based on the scaling relations (\ref{errn}), (\ref{viscexp}),
    (\ref{stressexp}) and (\ref{psiexp}). Those for Zimm dynamics are based on
    the data analysis of Fig.~\ref{fig:3}.  The values are listed for cluster
    statistics according to three-dimensional bond percolation (3D) and
    Erd\H{o}s--R\'enyi random graphs (ER), and are compared to some
    experimental findings.}
    \br
               & \centre{2}{Zimm} & \centre{2}{Rouse} &&& &&&&&&&&\\ \ns\ns
               & \crule{2} & \crule{2} &&& &&&&&&&&\\
      Exponent & 3D & ER & 3D & ER  & \cite{TaUr90} & \cite{DeBo93} &
               \cite{AxKo90} & \cite{AdLa97}& \cite{ZhZh96} & \cite{TaYo94} &
               \cite{LuMo95} & \cite{MaAd88}& \cite{CoGi93} & \cite{LuMo99} &
               \cite{TiTo04}      \\ \mr \lineup 
      $k$        & $^{(*)}$&$^{(*)}$& 0.71 & 0$^{\,(\#)}$ & 0.2 & 0.7 & 0.82 &
    1.1 & 1.27 & 1.3 & 1.36 & 1.4 & $>$1.4 & 6.1 &  \\
      $\ell$     &  1.3  & 0.54 & 4.1  & 3   &   &         &  &&&&&&&& \\
      $\Delta$   &       &      & 0.79 & 1   &   & 0.72 & 0.71 & 0.69 & &
    0.67 -- 0.68 & 0.66 & 0.70 & & 0.33 & 0.69 -- 0.77 \\
      $z$        &       &      & 3.3  & 3   &   & 2.9  & 2.67 &&&&&&&& \\ 
      $b_{\eta}$ &  0.11 & 0.11 & 0.50 & 1/2 &  &  &  & & & & & & & & \\
      $b_{\Psi}$ &  0.77 & 0.77 & 2.0  & 2   &  &  &  & & & & & & & & \\
    \br
  \end{tabular*}
  $^{(*)}$ no divergence \qquad $^{(\#)}$ logarithmic divergence
\end{table}

A fairly complete scaling picture of the gelation transition has been obtained
within Rouse dynamics. All critical exponents $k$, $\ell$, $\Delta$ and
$z$ of the stress-relaxation function in the sol
phase and at criticality could be expressed in terms of two independent static
percolation exponents $\sigma$ and $\tau$ plus the spectral dimension $d_{s}$
of the incipient percolating cluster, see the scaling relations
(\ref{viscexp}), (\ref{stressexp}) and (\ref{psiexp}). These scaling relations
and the resulting numerical exponent values listed in Table~\ref{stresstab}
contradict the predictions $k=2\nu -\beta$ and
$\Delta = d\nu /(d\nu +k)$ of earlier scaling arguments \cite{DuDe87,%
  MaAd88,CoGi93,Gen78,MaAd89,RuCo89}. What is the reason for this discrepancy?
The scaling arguments involve the fractal Hausdorff dimension $d_{f} := d
-\beta/\nu$ of \emph{rigid} percolation clusters at $c_{\mathrm{crit}}$. Rouse
clusters, however, are thermally stabilised, Gaussian phantom clusters with
the fractal Hausdorff dimension $d_{f}^{(\mathrm{G})}$, see (\ref{gausshaus})
\cite{Cat84,Vil88,SoBl95}. The latter is different from $d_{f}$ in space
dimensions below the upper critical dimension $d_{u}=6$. Indeed, if one
replaces $d_{f}$ by $d_{f}^{(\mathrm{G})}$ in these scaling arguments, as one
should consistently do within a Rouse description, the results will coincide
with the ones obtained here.

Since the long-standing scaling relations $k=2\nu -\beta$ and $\Delta = d\nu
/(d\nu +k)$ involve the Hausdorff fractal dimension $d_{f}$ of rigid
percolation clusters, it is sometimes argued that they describe the behaviour
of a more realistic model, which, in addition to the interactions of the Rouse
model, accounts for excluded-volume effects, too, see e.g.  \cite{RuCo89}. As
far as we know, this claim has not been verified by analytical arguments
within a microscopic model. One may even have doubts whether this claim is
generally true: Extensive molecular-dynamics simulations \cite{VePl01} of a
system of crosslinked soft spheres in three dimensions, with cluster
statistics from percolation and an additional strongly repulsive interaction
at short distances, yield the values $k\approx 0.7$ and $\Delta\approx 0.75$,
which are remarkably close to the predictions of the Rouse model for randomly
crosslinked monomers, see Table~\ref{stresstab}. On the other hand,
simulations of the bond-fluctuation model in \cite{GaAr00} imply $k\approx
1.3$ and are thus in favour of the claim. However, the viscosity is not
measured directly in these latter simulations. Rather it is derived from the
scaling of diffusion constants and an additional scaling assumption that may
be questioned \cite{VePl01}. Hence, it is an open problem to what extent the
critical Rouse exponents of Table~\ref{stresstab} are modified by
excluded-volume interactions.

In the context of dynamical critical phenomena, one usually expects dynamical
scaling to hold. Thereby one can infer critical properties of the gel phase
from those of the sol phase. In particular, the critical behaviour of the
shear modulus $ G_{0}\sim |\varepsilon|^{\mu}$ follows from the scaling form
(\ref{gscale}) of the stress-relaxation function. The result $\mu = \Delta z =
(\tau -1)/\sigma$ involves only the two exponents $\sigma$ and $\tau$ of the
cluster-size distribution. Using well-known scaling relations of percolation
theory, this can be rewritten as $\mu = d\nu$ in terms of the
correlation-length exponent $\nu$ and the spatial dimension $d$. It is in
agreement with the simple scaling argument based on dimensional analysis of
the free-energy density. In a recent letter \cite{XiMu04}, the scaling of
entropic shear rigidity was analysed for both phantom chains and those with
excluded-volume interactions. In both cases the gel was prepared by
crosslinking a melt of chains with excluded-volume interactions. Our choice of
percolation statistics combined with Rouse dynamics should be comparable to
phantom chains prepared in an ensemble with excluded-volume interactions.
However, the results of \cite{XiMu04} for $\mu$ disagree with the above
dynamic-scaling argument. The reasons for the discrepancy are not understood.

Let us return to the sol phase and discuss the Zimm results, which are based
on the exact numerical determination of the scaling exponents $b_{\eta}$ and
$b_{\Psi}$ for the fixed-size averages (\ref{zimmscaling}) of the viscosity
and of the first normal stress coefficient. The resulting finiteness of the
macroscopic viscosity $\langle\eta\rangle$ at the transition is clearly the
most serious drawback of the Zimm model for randomly crosslinked monomers.
Also, this failure comes unexpected, because a well-known scaling argument
\cite{StCo82,Mut85,Cat84} predicts a logarithmic divergence. This scaling
argument uses the (correct) scaling $D_{n}\sim 1/R_{\mathrm{gyr},n}$ of
diffusion constants together with the Stokes--Einstein relation and yields
$b_{\eta}=d/d_{f}-1$. Consequently, one gets from (\ref{critscaling}) the
scaling relation $k=(1-\tau+d/d_f)/\sigma$. Inserting hyperscaling and the
fractal dimension of rigid percolation clusters, one would get $k=0$ from
that, which was interpreted as a logarithmic divergence.  But as we remarked
already earlier on in this subsection, the correct fractal dimension $d_{f}$
for Gaussian phantom clusters is $d_{f}^{(\mathrm{G})}$. For both cluster
statistics this would give an unphysical negative value around $-1/4$ for
$b_{\eta}$ which can be definitely ruled out by our
data.\footnote{Unfortunately, the value of $b_{\eta}$ resulting from this
  scaling argument in the case of Erd\H{o}s--R\'enyi random graphs was
  incorrectly ascribed to $d=6$ dimensions in the second last paragraph of
  \cite{LoMu04}, leading to the wrong statement $b_{\eta} = 1/2$ there.} Thus,
we conclude that the scaling approach of \cite{StCo82,Mut85,Cat84} does not
apply to the Zimm model for randomly crosslinked monomers. Another scaling
approach to this model by \cite{ArSa90} is also falsified by our data. On the
other hand, Brownian-dynamics simulations of hyperbranched polymers were
performed in \cite{ShAd02}. They also account for \emph{fluctuating}
hydrodynamic interactions corresponding to $\kappa = 0.35$, as well as for
excluded-volume interactions and lead to $b_{\eta} = 0.13$.  This result is
remarkably close to our finding $b_{\eta} \approx 0.11$ for the highest
coupling strength $\kappa =0.3$ that we have considered, whereas experimental
findings (see below) are consistently above our value.

Next, we comment on how the Rouse and Zimm predictions for stress relaxation
compare to experimental reality.  Table~\ref{stresstab} shows an enormous
scatter of the experimental data.  Thus, a serious check of theoretical
predictions is currently severely hampered. The origin of this wide spread of
the data is unclear so that even the question arose, whether the dynamical
critical behaviour at the gelation transition was indeed universal
\cite{AdDe85}.  Possible explanations for non-universal behaviour include the
splitting of a static universality class into two dynamical ones
\cite{DaLa90,ArSa90} and, for the case of crosslinking long polymer-chain
molecules (vulcanisation), a decrease of the width of the critical region with
increasing chain length \cite{Gen77}.  The latter may explain the observation
of a crossover behaviour to mean-field properties in certain gelation
experiments, if measurements were not performed well inside the true critical
region.

As far as we know, no measurements of the critical behaviour of the first
normal stress coefficient have been reported.  The Rouse value $k \approx
0.71$ for the viscosity and three-dimensional percolation statistics agrees
well with the experiments of \cite{AdDe81,AdDe85,DuDe87,DeBo93} (only
\cite{DeBo93} was included in Table~\ref{stresstab} to demonstrate the broad
scatter of the viscosity data). On the other hand, it is not compatible with
the possibly oversimplifying albeit attractive proposal \cite{DaLa90,ArSa90}
to interpret the wide variation of the viscosity exponent $k$ as a signature
of a splitting of the static universality class of gelation into different
dynamic ones. Indeed, Rouse and Zimm dynamics are considered
\cite{Oon85,DoEd88} to be at the extreme ends of the strength of the
hydrodynamic interaction. Since the Zimm model does not even predict a
divergence at the transition, the actual value of $k$ should then lie below
the Rouse value according to that proposal.

Hence, the broad scatter of the experimental data calls for additional
relevant interactions beyond those accounted for in the Zimm or Rouse model.
This may be due to the preaveraging approximation. In particular, it throws
away hydrodynamic interactions among different clusters.  But we do not expect
this to be the sole relevant simplification of the Zimm model, because linear
polymers show a decrease in the viscosity when abandoning the preaveraging
approximation \cite{Fix81}, and effects of preaveraging for branched molecules
are even more pronounced than those for linear ones \cite{BuSc80}. Rather it
seems that there are no satisfactory explanations without considering
excluded-volume interactions.  Indeed, simulations \cite{GaAr00} of the
bond-fluctuation model deliver higher values $k\approx 1.3$ in accordance with
the scaling relation $k=2\nu -\beta$, which arises from heuristically merging
Rouse-type and excluded-volume properties, see above. On the other hand,
entanglement effects are neglected, too.  These topological interactions are
argued to play a vital role in stress relaxation. However, \emph{temporary}
entanglements are expected to play only a minor role \cite{CoGi93} for the
dynamics close to the gelation transition. This is because the time scale of a
temporary entanglement is determined by the smaller clusters, whereas
near-critical dynamics is determined by the largest clusters, which contribute
the longest time scales. Yet, there remain \emph{permanent} entanglements due
to interlocking loops. They are clearly far beyond the scope of the present
and many other theoretical approaches.

%
\section{Closing remarks}
\label{closing}
%

The list of shortcomings of the Rouse and the Zimm model for crosslinked
monomers is long, and it was discussed in Sections \ref{densitydiscuss} and
\ref{stressdiscuss}.  Yet, one should not underestimate the importance of
these models for our understanding of gelling liquids. 
First, the success of Theoretical Physics and, in particular, Statistical
Physics has always relied on capturing the essence of observable phenomena in
simple mathematical models. Models that isolate certain physical mechanisms
and, at the same time, sacrifice many details of the observed reality. It is
safe to say that, at least for linear polymers, the Rouse and the Zimm model
have proven to be among this class \cite{BiCu87,DoEd88}.  Second, simple
exactly solvable models always represent cornerstones against which more
elaborate theories, approximation methods and numerical simulations can be
tested. Moreover, in the absence of an ultimate theoretical picture, the
predictions of such minimal models also serve as a standard reference for
experimental data.  Indeed, this has been common practice in experimental
investigations on gelling liquids over the years, see e.g.\ the review
articles \cite{MaAd91,WiMo97}. All the more it is important to have reliable
and mathematically firmly based predictions of these model.

\ack{ 
  We gratefully acknowledge the collaboration with M.\ K\"untzel on
  self-diffusion in the Zimm model.  This work was supported by the Deutsche
  Forschungsgemeinschaft (DFG) under grant numbers Zi~209/6-1, Zi~209/7-1,
  Mu~1056/2-1 and through SFB 602.  
}

%
\section*{References}
%

\end{document}